\documentclass[9pt,twocolumn,twoside]{pnas-new}

\templatetype{pnasresearcharticle}
\usepackage{graphicx,amssymb,amsmath,epsf,bm,cprotect,comment,physics,titlesec,xcolor,array,bbold}

\epsfclipon

\newcommand{\ri}{\mathrm{i}}
\newcommand{\re}{\mathrm{e}}

\newcommand{\unit}[1]{\,\mathrm{#1}}

\newcommand{\rd}{\mathrm{d}}
\newcommand{\expct}[1]{\langle{#1}\rangle}
\newcommand{\Expct}[1]{\left\langle{#1}\right\rangle}
\newcommand{\diff}[2]{\frac{\mathrm{d} #1}{\mathrm{d} #2}}

\renewcommand{\eqref}[1]{Eq.\,[\ref{#1}]}

\newcommand{\figref}[1]{Fig.\,\ref{#1}}
\newcommand{\figpref}[2]{Fig.\,\ref{#1}\textit{#2}}

\newcommand{\supfigref}[1]{SI Appendix, Fig.\,\ref{#1}}
\newcommand{\supfigpref}[2]{SI Appendix, Fig.\,\ref{#1}\textit{#2}}
\newcommand{\suptblref}[1]{SI Appendix, Table\,\ref{#1}}
\newcommand{\vidref}[1]{Movie\,S#1}

\usepackage{xr} 
\makeatletter
\newcommand*{\addFileDependency}[1]{
  \typeout{(#1)}
  \@addtofilelist{#1}
  \IfFileExists{#1}{}{\typeout{No file #1.}}
}
\makeatother

\newcommand*{\myexternaldocument}[1]{
    \externaldocument[S-]{build/#1}  
    \addFileDependency{#1.tex}
    \addFileDependency{build/#1.aux}  
}
\myexternaldocument{suppl-arxiv}  

\begin{document}

\title{Rigidity transition of a highly compressible granular medium}

\author[a,1,2]{Samuel Poincloux}
\author[a,b]{Kazumasa A. Takeuchi}

\affil[a]{Department of Physics,\! The University of Tokyo,\! 7-3-1 Hongo,\! Bunkyo-ku,\! Tokyo 113-0033,\! Japan}
\affil[b]{Institute for Physics of Intelligence,\! The University of Tokyo,\! 7-3-1 Hongo,\! Bunkyo-ku,\! Tokyo 113-0033,\! Japan}

\leadauthor{Poincloux}

\significancestatement{When particles in disordered materials are densely packed, they undergo a jamming transition, switching from a flowing assembly to a solid. Whereas the state-of-the-art understanding of this transition -- in terms of constraints given by the number of particles’ contacts -- has been established for hard or moderately soft particles, materials made of extremely deformable particles are also common in biological, geological, and engineering fields. Here, we introduce a model experimental system made of such “squishy” particles, a discrete assembly of highly compressible but mobile particles. We discover a characteristic rigidity transition controlled by the size of contacts instead of their number, redefining the concepts of constraints. Our study highlights a geometry-friction interplay underlying this novel transition.}

\authorcontributions{S.P. and K.A.T. designed research; S.P. performed research; S.P. and K.A.T. contributed new reagents/analytic tools; S.P. and K.A.T. analyzed data; and S.P. and K.A.T. wrote the paper.}
\authordeclaration{The authors declare no competing interests.}
\equalauthors{\textsuperscript{1}Current affiliation: Department of Physical Sciences, Aoyama Gakuin University, 5-10-1 Fuchinobe, Sagamihara, Kanagawa 252-5258, Japan}
\correspondingauthor{\textsuperscript{2}To whom correspondence should be addressed. E-mail: poincloux@phys.aoyama.ac.jp}


\begin{abstract}
A wide range of disordered materials, from biological to geological assemblies, feature discrete elements undergoing large shape changes. How significant geometrical variations at the microscopic scale affect the response of the assembly, in particular rigidity transitions, is an ongoing challenge in soft matter physics. However, the lack of a model granular-like experimental system featuring large and versatile particle deformability impedes advances. Here, we explore the oscillatory shear response of a sponge-like granular assembly composed of highly compressible elastic rings. We highlight a progressive rigidity transition, switching from a yielded phase to a solid one by increasing density or decreasing shear amplitude. The rearranging yielded state consists of crystal clusters separated by melted regions; in contrast, the solid state remains amorphous and absorbs all imposed shear elastically. We rationalize this transition by uncovering an effective, attractive shear force between rings that emerges from a friction-geometry interplay. If friction is sufficiently high, the extent of the contacts between rings, captured analytically by elementary geometry, controls the rigidity transition.
\end{abstract}

\dates{This manuscript was compiled on \today}
\doi{\url{www.pnas.org/cgi/doi/10.1073/pnas.XXXXXXXXXX}}

\maketitle
\thispagestyle{firststyle}
\ifthenelse{\boolean{shortarticle}}{\ifthenelse{\boolean{singlecolumn}}{\abscontentformatted}{\abscontent}}{}


Particulate assemblies, such as granular media, can switch their mechanical behavior between a solid and a fluidized response depending on their density, applied stress or interactions. 
This rigidity transition is of tremendous importance across length scales and disciplines, with the elemental particles varying in size, shape, and stiffness. 
Such jamming or yielding transitions are, for instance, underlying the onset of geological flows \cite{kostynick_rheology_2022}, the transport and clogging in industrial food processing \cite{xu_modelling_2018}, and are at the heart of numerous biological processes \cite{hannezo_rigidity_2022}. 

For athermal rigid particles without internal degrees of freedom, the rigidity transition, called jamming, is mainly controlled by the average number of constraints imposed by the number of neighbors \cite{henkes_critical_2010}, including for non-spherical particles \cite{brito_universality_2018}.
Even when particles are softer and can moderately deform their shape upon contact, such as emulsions or foams, the number of contacts remains an essential factor \cite{brujic_measuring_2007,winkelmann_2d_2017,boromand_role_2019}. 
However, particle deformations may reach up to the order of the particle size for a broader class of materials, from biological tissues to porous soils.
By adapting their shape to external constraints, such "squishy" particles \cite{bares_softer_2022} acquire additional degrees of freedom, potentially affecting the ingredients underlying the jamming transition.

Recent works highlight the role of shape and deformability in the jamming transition of highly deformable particles, both in thermal \cite{gnan_microscopic_2019,gnan_dynamical_2021}  and athermal \cite{boromand_jamming_2018,wang_structural_2021,treado_bridging_2021,clemmer_soft_2024} assemblies.
Dedicated tools have also been developed to quantify the large shape changes that cells undergo while embedded in a deforming tissue \cite{guirao_unified_2015}. 
The shape of the cells in epithelial tissues is one of the essential parameters governing a rigidity transition \cite{bi_density-independent_2015,park_unjamming_2015,bi_motility-driven_2016,arora_shape_2024}. 
In geological flows, the shape of constitutive particles undergoes significant changes primarily via breaking \cite{seo_evolution_2021}, which in turn have a critical effect on the rheology \cite{Hu_weakening_2020} and jamming transition \cite{zhang_grain_2016,kostynick_rheology_2022}. 
With very soft grains that do not break, compaction far above jamming highlights that not only the coordination number increases with compaction, but also the size of the contacts \cite{mukhopadhyay_packings_2011,bares_compacting_2023}.

Despite these recent advances, a clear picture of how extreme deformability affects the jamming transition is still lacking. 
It is an ongoing challenge in soft matter physics and related fields \cite{manning_essay_2023}. 
Although advanced numerical methods, such as the vertex model \cite{bi_motility-driven_2016}, the material point method \cite{nezamabadi_implicit_2015} and the phase field approach \cite{negro_yield-stress_2023,saito_cell_2024}, have been used, numerical investigations remain incredibly challenging owing to the non-linear couplings between geometry and interactions at the local scale, and a large number of elements.
As such, a simple experimental model featuring large deformation at the grain scale would be essential to explore this question. One may leverage the recent advances in granular metamaterials \cite{haver_elasticity_2024} and metafluids \cite{djellouli_shell_2024}, whose properties are governed by geometrical non-linearities at the particle scale, to control grains compressibility.

Here, we introduce assemblies of compressible elastic rings as an experimental model to explore the role of large grain deformations in the rheology of disordered media. 
Via oscillatory shear experiments, we highlight a rigidity transition controlled by the shear amplitude and the density.
While the transition is characterized by the absence or presence of structural rearrangements, compression-induced large geometrical deformations help the assembly absorb shear without rearranging in the solid phase.
Finally, we argue that this transition is mainly driven by the size of the frictional contacts between the compressed rings via effective adhesive interactions under shear, which stem from the interplay between geometrical deformations and friction.

\section*{Rigidity transition in a sheared assembly of compressible rings}

\begin{figure}[t]
\centering
\includegraphics[width=\hsize]{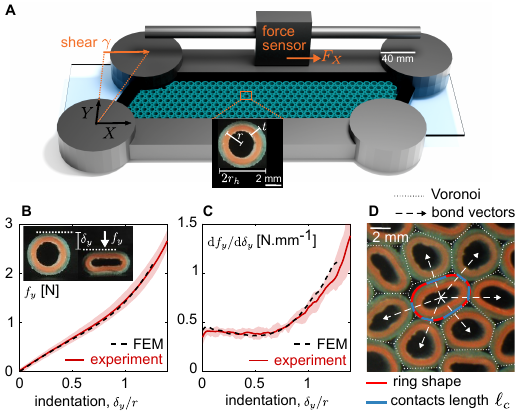}
\caption{Oscillatory shear of an assembly of compressible rings. 
(A) Schematic of the experimental setup. A ring assembly is confined within a 2D cell with a fixed area. Oscillatory shear $\gamma=(\gamma_0/2)\sin \left(\dot{2\gamma_0}t/\gamma_0\right)$ is applied to the assembly via lateral motion of the top arm while the bottom one is fixed. Both top and bottom arms are decorated with half-solid disks of the same dimension as the rings to transmit shear. A transparent acrylic plate placed on top keeps the rings in the plane.
The inset shows an image of a single ring.
The thin orange silicone layer is applied in the inner part of the ring to help with detection and further imaging processes. 
(B,C) Experimental and simulated (FEM) indentation response $f_y(\delta_y)$ (B) and associated stiffness $\diff{f_y}{\delta_y}$ (C) of a single ring compressed between two flat plates. The experimental curve is the averaged response over five different rings from the assembly; upon loading and unloading, the shaded area shows the minimal and maximal values reached by the rings during the compression cycle. The corresponding FEM response is made via a 3D simulation of the neo-Hookean material model with Young's modulus $E=1\unit{MPa}$ and Poisson's ratio $\nu=0.45$, reproducing the ring and indentation geometry. The experiments and FEM, which agree quantitatively, feature a slight softening upon indentation, leading to solid foam-like deformation heterogeneities upon compression of the ring assembly. 
(D) For each ring, we measure via image analysis its position (geometric center), shape, neighbors, and extent of each contact (length of the blue segments). We define two neighborhood types: Voronoi neighbors share a common edge in the Voronoi tessellation, and touching neighbors are in contact with a finite contact extent $\ell_c$. The bond vectors between touching neighbors are represented by dashed arrows. See also \vidref{2}. 
}
\label{fig1-1}
\end{figure}

We consider a two-dimensional assembly of frictional, elastic, and ring-shaped grains (see \figpref{fig1-1}{A} and \vidref{1}). 
To explore the rheological response of the ring assembly, we use a typical oscillatory shear method \cite{marty_subdiffusion_2005,kawasaki_macroscopic_2016,otsuki_shear_2020,das_unified_2020,otsuki_softening_2022} where the assembly is compressed into an imposed area and sheared by oscillatory displacement of the top boundary as shown in the setup schematic in \figpref{fig1-1}{A}. 
The rings are all identical with a thickness $t=1.5\unit{mm}$, radius $r=3.3\unit{mm}$ and height $h=10\unit{mm}$.
Rings are fabricated via molding of polymerizing liquid silicone (see Methods and \supfigref{S-fig_sup_ring}), then coated with talc powder to avoid adhesion and ensure dry frictional interactions.
The compression response of the rings to a loading-unloading cycle (\figpref{fig1-1}{B}) attests to a perfect elasticity with no remaining permanent deformations and high reproducibility, and is well recovered by the finite element method (FEM) (see Methods for details). 
In addition, the onset of the compression response features a small stiffness decrease (\figpref{fig1-1}{C}). 
Deformed rings being slightly softer, a compressed assembly of such rings develops heterogeneous deformations as shown in the top view displayed in \figpref{fig1-1}{D} (see also \vidref{1}). 
This heterogeneous deformation upon compression is reminiscent of the mechanics of cellular materials \cite{poirier_experimental_1992,papka_biaxial_1999,combescure_post-bifurcation_2016}, including solid foams \cite{yang_crushing_2020}, with the critical difference that here the unit cells do not adhere each other and can therefore rearrange. 

The assembly's response to shear is explored by systematically varying the shear amplitude $\gamma_0$ and the density of rings $\Phi$. 
The oscillatory shear $\gamma$ is imposed at a fixed maximum shear rate of $\dot{\gamma_0}=1.33\times10^{-2}\unit{s^{-1}}$, low enough to ensure a quasi-static regime where inertial or vibrational effects are negligible.
The density is adjusted by varying the number of rings in the shear cell and quantified by $\Phi=N\pi r_h^2/A$ with $r_h=r+t/2$ the external radius of the rings, $N$ the number of rings and $A$ the area of the shear cell. 
This is the density that would correspond to the undeformed area of the rings, and we focus on the regime above the hexagonal packing density $\Phi_h=\frac{\pi}{2\sqrt{3}}\approx0.91$. 
Initialization is made by dispersing the rings in the cell area and then manually lowering and fixing the top arm at the desired height. 
The ring assembly is, therefore, initially in a mechanically quenched state. 
An acrylic plate on top of the assembly holds the rings in 2D, preventing out-of-plane buckling \cite{ghosh_popping_2024}. 

\begin{figure}[t]
\centering
\includegraphics[width=\hsize]{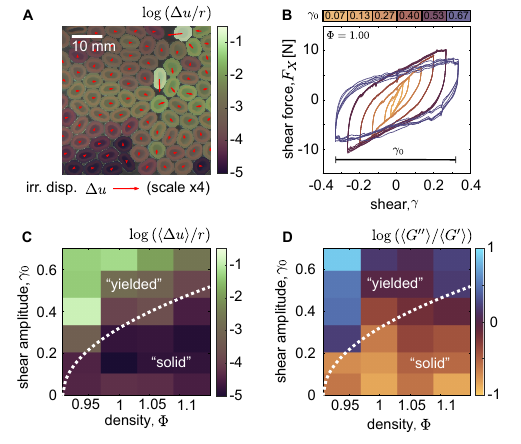}
\caption{Rigidity transition in the density-shear amplitude phase space. 
(A) Top view of a portion of the ring assembly ($\Phi=1.04$ and $\gamma_0=0.53$).
Here, each tracked ring is colored over its detected area by the value of its irreversible displacement $\log(\Delta u/r)$. The corresponding displacement is shown by a red arrow (up-scaled by a factor $4$). 
(B) Shear response $F_X(\gamma)$ of the assembly for $\Phi=1.00$ and for shear amplitude $\gamma_0$ ranging from $0.07$ to $0.7$. Six cycles in the steady state are displayed for each $\gamma_0$. 
(C) Intensity of the averaged irreversible displacement $\log(\Expct{\Delta u}/r)$ for varying density $\Phi$ and shear amplitude $\gamma_0$. The average is calculated over all tracked rings and in the steady state. 
(D) Map of the ratio between the loss and storage moduli of the assembly, computed from the shear response $F_x(\gamma)$ and averaged over the cycles in the steady state. The corresponding $F_x(\gamma)$ for $\Phi=1.00$ are shown in (B), each colored by its corresponding value of $\log(\langle G''\rangle/\langle G'\rangle)$. In ($C$ and $D$) and subsequent maps, the white dotted line is a geometric prediction introduced later (\eqref{eq1}), which serves here as a reference to help compare the maps and identify solid and yielded regions. 
}
\label{fig1-2}
\end{figure}

The mechanical response to shear is probed locally via image analysis and globally via a force sensor. 
Pictures from an overhanging camera are taken regularly during the cycles, such that the position, shape and neighbors of each ring are identified and tracked all along the shear cycles (\figpref{fig1-1}{D}; see also Methods and \vidref{2}). 
Image analysis permits to measure quantities such as the irreversible displacement between two consecutive cycles (\figpref{fig1-2}{A}).   
A force sensor, placed between the driving screw and the top arm, records the force $F_X$ necessary to shear the assembly. 
The shear force $F_X$ for $\Phi=1.0$ and all investigated shear amplitudes is shown in \figpref{fig1-2}{B}. 
Unless stated otherwise, the presented measures represent the system in the steady state (see \suptblref{S-suptab1} for the cycles used in the analysis). 
The high reproducibility of the shear response in \figpref{fig1-2}{B} testifies that the assembly indeed reached a steady state and that results are robust against which cycles are considered in the steady state (\supfigref{S-fig_sup_same_cycle}).
We also verified that in the steady state, the experiment is reproducible and not sensitive to the initial configuration (\supfigref{S-fig_sup_repro}). 

We then assess whether the ring assembly responds more in a solid-like or yielded fashion.
To assess this response, a classic measure for particulate systems under oscillatory loading is the irreversible displacement of the particles between two consecutive cycles \cite{keim_mechanical_2014,dagois-bohy_softening_2017,nagasawa_classification_2019,otsuki_shear_2020,das_unified_2020,otsuki_softening_2022,zeng_equivalence_2022}. 
We define $\Delta u(i,c)$ as the norm of the irreversible displacement of ring $i$ occurring between the start and end of cycle $c$. In \figpref{fig1-2}{A}, rings are colored by their value of $\log\left(\Delta u /r\right)$ and the associated irreversible displacement is drawn with a red arrow (magnified by a factor 4). 
Even at such a high density, irreversible displacements occur but are highly heterogeneous, with values spanning orders of magnitude within the assembly. 
To explore how the irreversibility depends on the density and shear amplitude, for each parameter pair $(\Phi,\gamma_0)$ considered, we compute $\langle \Delta u \rangle$, the averaged irreversible displacement over all tracked rings in the steady state.
The resulting map is shown in \figpref{fig1-2}{C}. 
Two zones can be identified: a yielded regime with significant irreversibility, which smoothly transitions to a solid-like zone with barely any irreversible displacement, by increasing density or decreasing shear (see also non-affine trajectories in \vidref{3}).

The transition between the yielded and solid-like features is observed not only in the rings' local movement but also via the global mechanical response. 
From the shear response $F_X(\gamma)$ of the whole assembly, we compute for each cycle the first harmonic of the storage and loss moduli, $G'$ and $G''$, respectively \cite{hyun_review_2011} (see separated map of $G''$ and $G'$ in \supfigref{S-fig_sup_moduli}). 
Identifying when the ratio $G''/G'$ is greater than $1$ is a standard method to evaluate if the assembly yields or remains essentially solid \cite{bonn_yield_2017}. 
The map of $\log(\Expct{G''}/\Expct{G'})$ shown in \figpref{fig1-2}{D}, with $\expct{G''}, \expct{G'}$ the moduli averaged over the cycles in the steady state, recovers qualitatively the solid and yielded regions identified above from the irreversible displacements (\figpref{fig1-2}{C}). 
Indeed, the region with high irreversible displacements coincides with a loss modulus significantly larger than the storage one and inversely for the dominantly reversible region. 

Exploring the mechanical response to shear at very high density and shear amplitude, we identify a progressive rigidity transition of the ring assembly by either reducing the shear amplitude or increasing the density. 
To understand the driving factors of this transition, we first quantify the structural properties of the assembly.

\section*{A partially crystallized yielded assembly and an amorphous solid}

The presence or absence of irreversible displacements upon cyclic shear in the steady state highlights that ring rearrangements and associated structural deformations play a crucial role in the mechanical response of the assembly \cite{richard_predicting_2020,galloway_relationships_2022}. 
We then measure the structural differences between the solid and yielded regions and how they evolve from their initial quenched state. 
With all rings having the same size, the configuration with the highest density and minimal shape deformation (and hence minimum elastic energy) is the hexagonal packing. 
We thus use the hexagonal bond orientational order $\Psi_6$ as the structural indicator of the assembly \cite{schreck_tuning_2011}. 
For each ring $j$, the hexagonal order is computed by $\Psi_6(j)=(1/n_j)|{\sum_{k=1}^{n_j}\re^{6\ri\alpha_{jk}}}|$ with $n_j$ the number of Voronoi neighbors and $\alpha_{jk}$ the angle between the line passing through the center of the two rings and the $x$ axis (the direction of the shear). 
The value of $\Psi_6$ is $1$ for a ring in a perfect hexagonal lattice and continuously decreases to $0$ as the structure deviates from the hexagonal one. 

\begin{figure}[t!]
\centering
\includegraphics[width=\hsize]{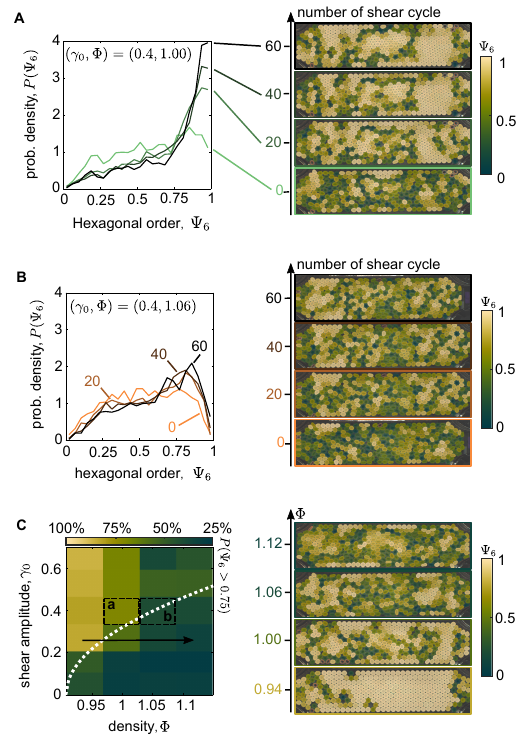}
\caption{Crystallizing yielded assembly and amorphous solid. (A) Left, probability density profiles of the hexagonal bond orientational order $\Psi_6$ for an assembly at $(\gamma_0,\Phi)=(0.4,1.00)$ in the yielded region. 
The profiles are shown for the assembly before shear and after $20$, $40$ and $60$ shear cycles. Right, corresponding snapshots of the assembly after cycling, where each ring is colored by its $\Psi_6$ value. 
The assembly shows a progressive growth of crystalline regions separated by melted ones.
(B) Same quantities as in (A) but for an assembly with higher density and same shear amplitude, $(\gamma_0,\Phi)=(0.4,1.04)$, now located in the solid region. Here, no significant structural changes are observed, and despite high shear, the assembly remains amorphous. See also \vidref{4}.
(C) Map of the proportion of rings belonging to crystal clusters (with $\Psi_6 > 0.75$) in the steady state. The corresponding snapshots of the assemblies in their steady states, with the rings colored by their $\Psi_6$ values, are shown on the right for $\gamma_0=0.25$ and varying densities. See also \vidref{5} for the time evolution of $\Psi_6$.}
\label{fig2}
\end{figure}

As the assembly accumulates shear cycles, we compare in \figref{fig2} (see also \vidref{4}) the structural evolution of configurations for two cases located near the border between the solid and yielded regions, with each case positioned on the opposite side of the border. 
After mechanical quenching, both configurations start in an amorphous state with a flat probability distribution $P(\Psi_6)$, but their evolution with cycling strongly differ. 
In the yielded region, for $(\gamma_0,\Phi)=(0.4,1.00)$ shown in \figpref{fig2}{A}, cycling induces drastic structural changes with the progressive growth of distinct crystal clusters.
The probability density increases close to $\Psi_6=1$ while it decreases for low values. 
In contrast, for larger densities within the rigid region (with $(\gamma_0,\Phi)=(0.4,1.06)$ shown in \figpref{fig2}{B}), no significant structural changes occur upon cycling, and the assembly remains in an amorphous phase. 
We then measure the crystallized portion of the assembly, or proportion of rings having $\Psi_6>0.75$, in the steady state over the parameter space (\figpref{fig2}{C}). 
All the assemblies in the solid region feature almost no structural changes and remain amorphous, while the proportion of the crystal clusters decreases with increasing density in the yielded region (\figpref{fig2}{C}; see also \vidref{5}). 
Rearranging rings within a given surface area tend to crystallize to reduce their overall elastic energy, as the hexagonal packing is the most compact structure without shape deformations. 
However, the growth of crystal clusters is limited by the presence of rings deformed by compression, which necessarily exist for high densities and prevent crystallization.

The structure of the ring assembly in its steady state differs drastically between the yielded and solid regions. The yielded region is characterized by hexagonal crystal clusters separated by a melted, disordered phase. This partial crystallization is typical of granular media under sufficient shear where reorganizations lead to structural changes \cite{panaitescu_nucleation_2012,hanifpour_mechanical_2014,royer_precisely_2015,rietz_nucleation_2018}. Even if essentially crystallized, the yielded phase still displays significant plastic flow with irreversible displacements (\figpref{fig1-2}{C} and $D$). In contrast, no crystallization occurs in the solid region, where the assembly remains amorphous almost without rearrangements. Even under high shear, the rings stick together to keep the same structure. In the following, we delve into the dynamics during shear cycles to investigate the interplay between structural and geometrical deformations of the rings. 

\section*{Large geometrical deformations absorb shear and hinder rearrangements}

\begin{figure}[t!]
\centering
\includegraphics[width=\hsize]{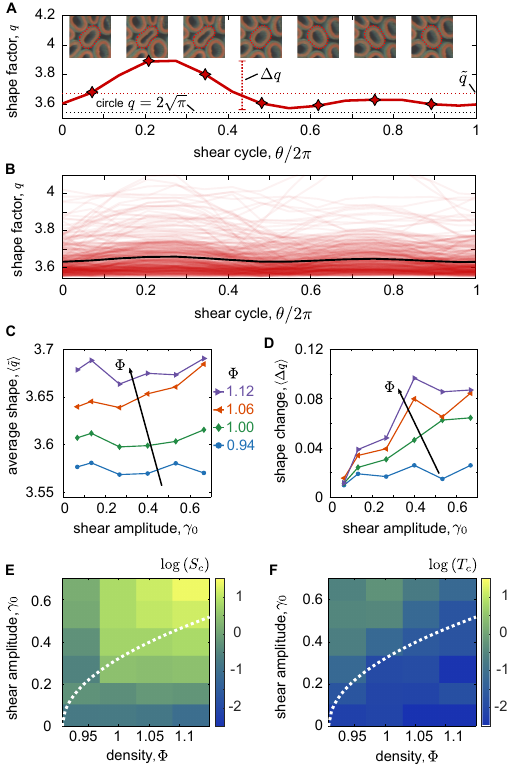}
\caption{Competition between geometrical and structural deformations.
(A) The geometrical deformation of the rings is quantified by the shape index $q$. The evolution of $q$ within a cycle is shown for one ring in a $(\gamma_0,\Phi)=(0.4,1.00)$ assembly. All along the shape trajectory, the shape of the ring at each diamond marker is shown in the snapshot above it. For each ring and each cycle, we extract the time-averaged shape index $\tilde q$ and the variation amplitude $\Delta q$ of its shape trajectory. 
(B) For the same assembly $(\gamma_0,\Phi)=(0.4,1.00)$, the shape trajectories of all the rings are represented by light red lines, while the average trajectory is marked by a black line. 
(C,D) The averages $\langle \tilde q \rangle$ (C) and $\langle \Delta q \rangle$ (D) evaluated over all tracked rings and in the steady state.
(E,F) Map of the geometrical $S_c$ and topological $T_c$ strains of the assembly. 
Strains are estimated using the method developed in \cite{guirao_unified_2015} (see also Methods) based on the frame-to-frame tracking of bond vectors between touching neighboring rings. 
The total strains here are the cumulative sums of the norm of the frame-to-frame strains estimated over the whole assembly and in the steady state. 
}
\label{fig3}
\end{figure}

We now focus on the deformation of rings during shear cycles to understand how the yielded and rigid regions respond to shear. 
The geometry of the rings is quantified by the shape index $q=(\mathrm{ring~perimeter})/\sqrt{\mathrm{ring~area}}$, a classic quantity used in highly deformable assemblies like epithelial tissues \cite{bi_density-independent_2015,park_unjamming_2015} to assess how far a convex shape is from a circle. 
The shape index takes the value $q=2\sqrt{\pi}\approx3.54$ for a perfect circle and increases when the ring gets deformed. 
Since the rings are slender objects, they deform mostly by flexion rather than stretching. Thus, $q$ is mainly increased by reducing the area while the perimeter is almost unchanged. 
This contrasts with emulsions or tissues where most deformations originate from a change in perimeter and not the area \cite{winkelmann_2d_2017,saito_cell_2024}. 
The perimeter and area of the rings are obtained via image analysis, and a Fourier decomposition is used to smooth the pixelated shape and reduce the shape to a few descriptors \cite{saito_cell_2024}. 
The resulting measure of $q$ during a shear cycle is illustrated in \figpref{fig3}{A} for one arbitrary ring within an assembly with parameters $(\gamma_0,\Phi)=(0.4,1.00)$. 
The diversity of shape trajectories for all the rings for a given cycle (\figpref{fig3}{B}) results from the heterogeneity of the assembly, but nonetheless the averaged trajectory is symmetric within a shear cycle.
For each cycle $c$ and tracked ring, we define two quantities from the shape trajectory: the average shape index $\tilde q=\Expct{q}_\mathrm{cycle}$ and the variation amplitude $\Delta q=\mathrm{max}(q)_\mathrm{cycle}-\mathrm{min}(q)_\mathrm{cycle}$ (\figpref{fig3}{A}), quantifying the mean deformation and the geometrical fluctuation, respectively.

In the steady state, the mean shape index and fluctuation are then averaged over all tracked rings, and their dependence on shear is compared for each density (\figpref{fig3}{C,D}). 
As expected when squeezing more rings in the same volume, the average geometrical deformation $\langle\tilde q\rangle$ increases steadily with the density (\figpref{fig3}{C}). 
While increasing shear amplitude has limited influence on $\langle\tilde q\rangle$, it drives significant shape fluctuations during shear cycles for all densities (\figpref{fig3}{D}). 
Moreover, the higher the density, the more pronounced the geometrical changes.
Thus, a higher density contributes, in addition to naturally increasing the average geometrical deformation, to promoting shape changes for a given imposed shear. 
Ring assemblies, therefore, use their compressibility and shape-changing abilities to absorb imposed external shear, and they do so more efficiently as the density increases. 

To grasp the combined role of geometrical changes and structural rearrangements leading to solid or yielded assemblies, we quantify these two aspects by dedicated strains used previously for epithelial tissue deformations \cite{guirao_unified_2015}, namely the geometrical strain $S^*$ and the topological strain $T^*$.
These can be evaluated from the dynamics of the bond vectors connecting touching neighbors (\figpref{fig1-1}{D} and \vidref{6}). 
While the precise definition is given in Methods (see also Ref.\cite{guirao_unified_2015}), essentially 
the geometrical strain $S^*$ is a contribution from the change in bond length and orientation and quantifies solely the continuous change in the size and shape of the bond network. 
In contrast, the topological strain $T^*$ is based on the creation and annihilation of bonds due to rearrangements and captures the discontinuous, topological change in the contact network.
The strains $S^*$ and $T^*$ are computed between each time frame and for the whole assembly. 
Their norm is then summed over all the cycles in the steady states to obtain the quantities $S_c=\int_t\norm{S^*}\rd t$ and $T_c=\int_t\norm{T^*}\rd t$ shown in \figpref{fig3}{E,F}, which quantify the degree of the geometrical deformation and the topological rearrangements, respectively, of the ring assembly over the shear cycles.
The map of $\log S_c$ in the parameter space $(\gamma_0,\Phi)$ (\figpref{fig3}{E}) reveals that geometrical changes increase significantly with both shear amplitude and density, regardless of the yielded or solid regions.
By sharp contrast, the map of $\log T_c$ in \figpref{fig3}{F} shows that topological deformations, while increasing with shear amplitude, are diminished for higher densities. 
As a result, the region with significant changes in the topology of the contact network coincides with the yielded region identified in \figpref{fig1-2}{C} and $D$. 
Therefore, while the yielded region features both topological rearrangements and geometrical deformations, the solid region is characterized by the quasi-absence of topological rearrangements, and consequently, the imposed shear is entirely absorbed by the geometrical deformations of the rings. 
It then appears that, for a fixed shear amplitude, an increase in density (and hence more deformed assemblies) hinders the structural rearrangements, leading to the solidification of the assembly. 
This result contrasts with confluent bi-dimensional active cell models, where more significant deformations lead to yielding \cite{bi_motility-driven_2016}. 
Now, to understand how large shape deformations impede structural ones in our ring assembly, we investigate the ring-ring interactions. 

\section*{Geometry and friction concur to generate effective shear adhesion between rings}

\begin{figure*}[t!]
\centering
\includegraphics[width=\hsize]{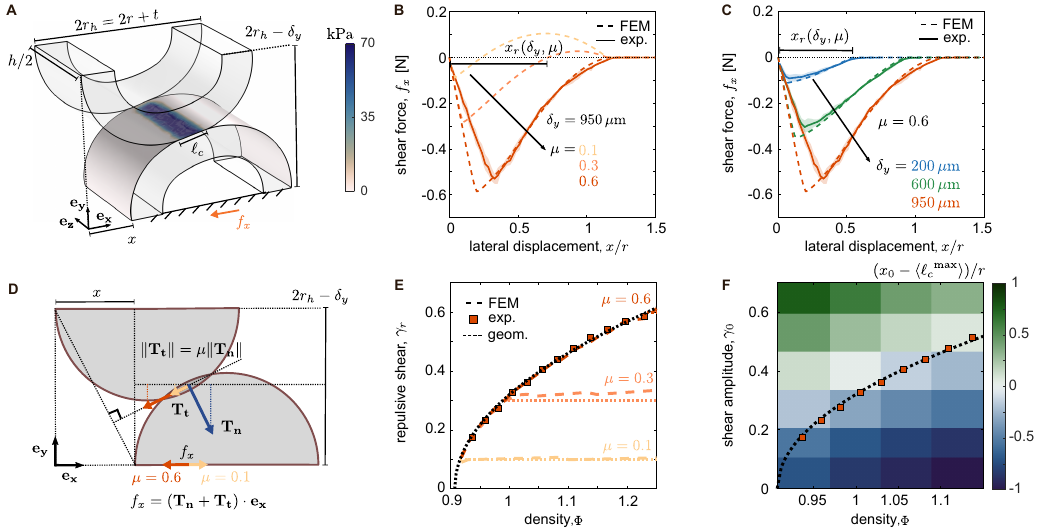}
\caption{Friction and geometrical deformations produce an effective shear adhesion between the rings.
(A) FEM simulations and experiments of ring-ring interactions. Two facing half-rings are compressed over a distance $\delta_y$ and then sheared laterally until they lose contact. The FEM snapshot shows the contact pressure between the rings after compression and a lateral shear displacement of $x$. See also \vidref{7}. 
(B,C) The resulting shear interaction force $f_x$ is measured as a function of the lateral displacement $x$. Negative (positive) values indicate an attractive (repulsive) force between the rings. Experimental responses are averaged over $3$ distinct pairs, and shaded areas represent max and min values obtained from the different pairs. 
Panel (B) shows the shear response $f_x(x)$ for $\delta_y=950\unit{\mu m}$ and the friction coefficient $\mu=0.1, 0.3, 0.6$ varied in the FEM. The experimental response matches that of FEM for $\mu=0.6$.
Panel (C) shows the shear response $f_x(x)$ for $\mu=0.6$, and increasing confinement $\delta_y=200, 600, 950\unit{\mu m}$. In this high friction regime, $f_x$ remains negative until the separation of the rings. We measure $x_r(\delta_x,\mu)$, the displacement needed to switch from attractive ($f_x<0$) to repulsive ($f_x > 0$) shear interactions. 
(D) Schematic of the ring-ring interaction on which the geometrical predictions are based. At a given shear distance $x$, the resulting shear force $f_x$ can either take positive or negative values depending on the friction coefficient (yellow and red arrows, respectively).  
(E) Experimental measure of the attractive-repulsive threshold shear $\gamma_r(\Phi)$, obtained by translating local quantities $(\delta_y,x_r)$ to global ones $(\Phi, \gamma_r)$, together with results from FEM (dashed lines) and the geometrical predictions (dotted lines) for $\mu=0.1, 0.3, 0.6$.
This curve $\gamma_r(\Phi)$ for $\mu=0.6$ is the one drawn in the phase diagrams shown so far (e.g., \figpref{fig1-2}{C} and $D$), marking the rigidity transition of the assembly (yielded for $\gamma_0>\gamma_r$ and solid-like otherwise).
(F) $(\Phi,\gamma_0)$ map of the difference between the imposed shear displacement $x_0$ and the measured maximal contact size $\langle{\ell_c^{\mathrm{max}}}\rangle$, averaged over all tracked rings and in the steady state. The geometrical prediction and local measurements of the attractive-repulsive threshold $\gamma_r$ correspond to the region where $x_0\approx\langle{\ell_{c}^\mathrm{max}}\rangle$. In this high friction and large geometrical deformations limit, the extent of the contacts $\ell_c$ mainly controls the rigidity transition.    
}
\label{fig4}
\end{figure*}

We introduce a microscopic ersatz comprising two half-rings sequentially compressed and sheared to measure the interaction between rings. 
Then, we will link the sign of the shear force, which would promote or prevent structural rearrangements, to the assembly's rigidity transition. 

We measure, via FEM and experiments, the shear response of two clamped rings compressed by an amount $\delta_y$ (\figpref{fig4}{A} and \vidref{7}). 
With FEM, we systematically vary the frictional interaction via the friction coefficient $\mu$ of the sheared interfaces (\figpref{fig4}{B}) and the compression $\delta_y$ between the rings (\figpref{fig4}{C}). 
Experimentally, we test rings taken from the assembly by applying the same compression values as in FEM. 
We find the best agreement between the experiments and the FEM for $\mu=0.6$, constituting a measure of the experimental friction coefficient. 
In the absence of friction, the shear force $f_x$ would be instantly positive (repulsive) as elastic deformations push the rings away from each other.
With friction, however, $f_x$ starts negative (attractive) and a finite lateral displacement $x_r(\delta_y,\mu)$ 
 is required to make it repulsive ($f_x > 0$) (see \figpref{fig4}{B}).
The synergy of geometrical deformation and friction thus produces an attractive interaction, or effective adhesion, between the rings until sufficient shear is applied. 
The role of shape change to generate an effective adhesion under shear is reminiscent of the rheology of cohesive granular media \cite{mandal_insights_2020}, where the low stiffness of the grains also participates in producing a higher effective adhesion. 
We argue next that the presence or absence of this effective adhesion under shear is at the root of the assembly's yielded or solid macroscopic response. 

We now give a geometrical prediction of the finite lateral displacement $x_r$ necessary to switch the shear interaction force from attractive to repulsive, i.e., $f_x(x_r)=0$. 
This distance is a function of the initial compression $\delta_y$ and the friction between the rings, $\mu$.
Two cases emerge in the way $x_r(\delta_y,\mu)$ is determined (\figpref{fig4}{D}).
In the first case, the rings remain in contact when $f_x$ switches from negative to positive (e.g., \figpref{fig4}{B} for $\mu=0.1$ and $0.3$). 
In this case, assuming Amontons-Coulomb frictional properties and reducing the contacting interfaces to a point, one can directly link the sign of the shear force $f_x$ to the lateral displacement and obtain $x_r(\delta_y,\mu)=\mu(2r_h-\delta_y)$.
By contrast, in the second case where $f_x$ remains negative until the rings are fully separated (e.g., \figpref{fig4}{C}), $x_r$ is given by the extent of the initial overlap between the rings: $x_r(\delta_y,\mu)=2r_h\sqrt{1-(1-\delta_y/(2r_h))^2}$. 
Now that we have a prediction for the microscopic ring-ring interaction, we must translate the compression $\delta_y$ and local lateral displacement $x$ to the global variable density $\Phi$ and shear $\gamma$. 
Assuming an hexagonal structure of density $\Phi_h$, an isotropic local overlap $\delta_y$ directly translates into a density by $\Phi(\delta_y)=\Phi_h/(1-\delta_y/(2r_h))^2$.
For the shear $\gamma$, too, by using its local counterpart, i.e., the local shear imposed between the two half rings, we obtain $\gamma = x/(2r_h-\delta_y)$.
Schematics of the underlying geometrical predictions are shown in \supfigref{S-fig_sup_schem}.
By combining the expressions of $x_r(\delta_y,\mu)$ with the expressions of $\Phi(\delta_y)$ and $\gamma(y,\delta_y)$, we can finally express the repulsive shear $\gamma_r$ at which the interactions become repulsive as a function of global variables $(\gamma,\Phi)$, as follows:
\begin{equation}
\gamma_r(\Phi,\mu)=
\begin{cases} 
      \sqrt{\frac{\Phi}{\Phi_h}-1} & \mathrm{if}\,\, \gamma_r < \mu \\
      \mu & \mathrm{if}\,\, \gamma_r \geq \mu
\end{cases}
\label{eq1}
\end{equation}
In \figpref{fig4}{E}, we compare $\gamma_r$ from the prediction against FEM and experimental measurements and find a perfect match.       
In the high friction case ($\gamma_r<\mu$), the threshold $\gamma_r(\Phi)$ is simply set by the extent of contacts, yielding $\gamma_r(\Phi)=\sqrt{\Phi/\Phi_h-1}$. 
Experimentally, the explored density range and the estimated friction coefficient $\mu\approx0.6$ are such that we remain in this high friction limit.
In soft spheres without shape changes, the jamming transition also features a scaling between yielding and density \cite{ikeda_unified_2012} but occurs at a much lower density and with a different scaling.

With repulsive interactions between rings, we expect structural rearrangements to be promoted and, inversely, hindered if interactions are attractive and act as an effective adhesive force. 
The shear $\gamma_r$ can then be viewed as a yield shear, where rearrangements leading to a yielded assembly are expected for $\gamma_0>\gamma_r$, and a solid-like assembly for $\gamma_0<\gamma_r$. 
Indeed, the line drawn by the geometric prediction of \eqref{eq1} qualitatively separates the yielded and solid-like regions in the $(\Phi,\gamma_0)$ parameter space (\figpref{fig1-2}{C} and $D$), which also accompany significant changes of the contact network topology (\figpref{fig3}{F}). 
This geometric boundary $\gamma_r(\Phi)=\sqrt{\Phi/\Phi_h-1}$ is set by the extent of the contact between rings, which increases with density. 
To verify that the contact extent is a major factor controlling the rigidity transition, we measure ${\ell_{c}^\mathrm{max}}$, the largest contact length among all touching neighbors, for each ring and at each time step.  
We then compute the difference between the equivalent local lateral displacement amplitude $x_0=\gamma_0 (2r_h-\delta_y)=\gamma_0 2r_h\sqrt{\Phi_h/\Phi}$ and $\langle{\ell_{c}^\mathrm{max}}\rangle$, averaged in the steady state (\figpref{fig4}{F}). 
The geometrical prediction of $\gamma_r$ matches with the region where $x_0\approx\langle{\ell_{c}}^{\mathrm{max}}\rangle$. 
The yielded and solid regions are then defined by whether the shear is large enough to overcome the effective adhesion and move the rings by a distance greater than the maximal contact length. 
Therefore, in contrast with rigid granular media where the number of contacts controls the jamming transition \cite{henkes_critical_2010}, the transition with squishy grains in a high friction limit seems to be strongly controlled by the extent of the contacts (\supfigref{S-fig_sup_coor}). 

\section*{Discussion}

We established that large shape deformation may radically modify the mechanisms by which an assembly yields. 
At high density, far into the solid state, the number of contacts becomes irrelevant, and their extent, coupled with frictional interactions, becomes crucial. 
The extra internal degrees of freedom provided by shape changes help the rings to resist shear without rearranging by accommodating their shape.
This is in sharp contrast with the classic picture of the jamming transition, where additional degrees of freedom would favor fluidization of the assembly, unless as many constraints are added to the system.
We argue that this contrast originates from geometry and friction concurring to generate an attractive force between rings under shear, where sufficient shear must be applied to overcome this effective adhesion.
In the high friction limit, the shear must be enough to separate the rings completely to overcome the adhesion; hence, a simple geometric formula (\eqref{eq1}) sets the transition between repulsive and attractive interactions. 
Below this transition (when the interactions are attractive), structural rearrangements are strongly hindered, then shape deformations take the upper hand and lead to a rigid assembly. 

However, these results are valid only within significant friction and shear limits. Our geometrical model implies that a frictionless assembly yields at an infinitesimal shear as $\gamma_r=\mu$. 
By considering only the interactions between two rings, we neglect collective effects, such as caging \cite{marty_subdiffusion_2005}, that would require a finite shear to yield even without friction. 
How caging is affected by significant shape changes is still an open question. 
Investigating the low and intermediary friction regime is then an important challenge left for future studies. This may be tackled if one can experimentally and systematically vary the friction coefficient between two deformable interfaces, for instance, by modifying the external texture of the rings \cite{maegawa_effect_2016}.

In addition, the nature of the observed rigidity transition is not fully elucidated. Further analysis of the particle dynamics via their mean-squared displacement in the steady state (\supfigref{S-fig_sup_transition}) hints that the transition shares traits of both a jamming and a glass transition, with the shear amplitude playing the role of an effective temperature \cite{ikeda_dynamic_2013}. However, the limited number of accessible cycles in the steady state hinders a conclusive characterization. 

Another striking property of the ring assembly is the drastic structure change between the yielded and solid states. 
Counter-intuitively, the yielded phase features highly crystallized portions, while the solid remains in an amorphous structure. 
Similarly to the link between active matter and sheared granular media \cite{morse_direct_2021}, one may also draw a parallel between this yielding crystal under shear and active crystals \cite{briand_spontaneously_2018,galliano_two-dimensional_2023}. 

From an engineering perspective, assemblies of highly deformable elements combine frictional interactions and structural rearrangements to dissipate energy while keeping a significant elastic component \cite{sano_randomly_2023}. 
Ring assemblies are then solid foams analogous with viscoelastic properties tunable by an external pressure field, with promising potential as a tunable shock absorber. A change in density around the rigidity transition point will greatly affect the material response to a shock.
To develop this axis, it would be essential to explore the 3D version of such assemblies and their response to inertial forces, in particular, how granular inertial flows \cite{jop_constitutive_2006} are affected by shape changes.

\matmethods{
\subsection*{Ring fabrication}
Rings are fabricated from silicone elastomer (Elite Double 22, Zhermack) in a urethane mold (Clear Flex 50, Smooth-On). 
The ring's material is chosen to remain elastic without permanent deformation, even under high and prolonged deformations. 
The mold material has been selected for its deformability, which eases demolding without tearing the rings, and its very low affinity with silicone, ensuring no adhesion with the rings and reusability. 
The rings have a thickness of $t=1.5\unit{mm}$, a midline radius $r=3.3\unit{mm}$ and height $h=10\unit{mm}$. 
To help with image analysis, a thin layer ($1\unit{mm}$) of dyed (Silc Pig Electric Orange, Smooth-On) silicone is applied on the inner portion of the ring surface (from $r-t/2$ to $r$). 
This extra layer does not affect the mechanics of the ring (the additional layer is not reproduced in FEM but still compares well with the experiments, \figpref{fig1-1}{B,C}). 
The rings are slightly coated with talc powder to avoid adhesion and ensure reproducible frictional properties. 
See \supfigref{S-fig_sup_ring} for an illustration of the protocol. 

\subsection*{Shear setup}
The shear cell comprises a fixed bottom arm and a movable top arm with movements constrained parallel to the bottom. 
The two sides of the cell freely rotate respectively to the top and bottom arms via joints of radius $40\unit{mm}$, and freely slide through the top arm via roller bearings to adjust their length under large oscillatory shear. 
The distance between the two side arms is $W=300\unit{mm}$, and the one between the top and bottom arm is $T=75\unit{mm}$. 
Both bottom and top arms are decorated by rigid (3D-printed in PLA and then glued) half disks of radius $r_h=r+t/2=4.05\unit{mm}$ (external radius of the rings). 
The whole setup rests on a fixed acrylic plate, and the ring assembly is confined vertically by a $3\unit{mm}$ thick acrylic plate blocked in the vertical direction but free to move laterally. 
The top arm imposes shear via a driving screw controlled by a stepper motor (AZM 46AC-TS10R, Oriental Motor). 
The motor imposes a cyclic horizontal motion $X=(X_0/2)\sin (2\dot{X_0}t/X_0)$, with $X_0$ the displacement amplitude and $\dot{X_0}=1\unit{mm/s}$ the maximum shear speed, kept constant for all the amplitudes considered. The resulting oscillatory shear is $\gamma=X/T$. 
The density of the rings is calculated by $\Phi=N\pi{r_h}^{2}/A$, where $A=16500\unit{mm^2}$ is the area of the shear cell, evaluated by taking into account the size of the joints and the top and bottom granular boundary conditions. 
With the highest shear considered, this area varies between $16500\unit{mm^2}$ and $17500\unit{mm^2}$ within a shear cycle ($\sim 6\%$ of variations). Consequently, densities are defined with uncertainties of $\sim \pm0.02$. 
The motor, force sensor, and camera are controlled and synchronized via a custom Labview program.

\subsection*{Force measurements}
All forces reported here are measured with a flat six-axis (forces and moments along the three directions) force sensor (USX10-H10-500N-C, TecGihan) and transferred to a computer via 2 data acquisition cards (USB-6001, National Instrument).
To measure the shear response of the assembly (\figpref{fig1-2}{B} and $D$), the sensor is inserted between the top arm and the driving screw.
The top arm is slightly lifted from the bottom acrylic plate to avoid measuring the frictional response of the setup. All six axes of the sensor are recorded at a rate of $20\unit{Hz}$. 
The first harmonic of the loss and storage moduli are computed by $G'=(2/\pi\gamma_0)\int_0^{2\pi}F_y(\theta)\sin\theta\rd\theta/Wh$ and $G''=(2/\pi\gamma_0)\int_0^{2\pi}F_y(\theta)\cos\theta\rd\theta/Wh$ with $\theta=(t-t_c) 2\dot{\gamma_0}/\gamma_0$ the normalized time since the start of cycle $c$ \cite{otsuki_softening_2022}. 
To measure the compression response of the rings (\figpref{fig1-1}{B,C}) and their interactions (\figpref{fig4}{B,C}), a distinct setup is used where positions are imposed manually via a micrometric stage and $100\unit{mm}$ long arm levers, and forces are measured via the moments output of the sensor. 
For the compression response, one free ring is confined between two rigid walls, while for the interaction experiments, a PLA 3D-printed part holding half of the rings into a groove clamps the pair of rings. 
For each interaction experiment, the six axes of the sensor are recorded continuously at $100\unit{Hz}$ while compression and shear are imposed via manually moving the micrometric stage. 
The displacement is made by step of $50\unit{\mu m}$ every $10\unit{s}$. 
For every step, the position is changed between $0$ and $5\unit{s}$, and the reported values for this new position are averaged over the time interval between $7$ and $9\unit{s}$. 
The arm lever's high but finite stiffness is measured in compression and shear and used to correct the displacement imposed on the rings. 

\subsection*{Finite Element Method (FEM)}
The mechanical tests using full 3D FEM were performed with the software COMSOL Multiphysics 6.1. 
We use symmetries to model only the necessary portions of the ring for the compression (\figpref{fig1-1}{B,C}) and shear (\figref{fig4}) simulations.
The rings are cylinders of height $h$ along the $z$ axis and of external radius $r_h$ in the $xy$ plane. 
Rings are compressed and sheared in the $xy$ plane, and we assume stress-free surfaces at $z=0$ and $z=h$. 
From symmetries, we model only half of the rings, from $z=0$ to $z=h/2$, with a free surface at $z=0$ and a mirror symmetry (no outward displacement) at $z=h/2$.
The tetrahedral mesh size is fixed and set to a minimum of $\sqrt{rt}/5$, chosen to optimize convergence.
The material is the Neo-Hookean model with Young modulus $E=1\unit{MPa}$ and Poisson's ratio $\nu=0.45$.
The Young modulus is determined by the best fit to the experimental data, while the Poisson ratio is taken just below $0.5$ as this silicone is nearly incompressible, but taking the limit value of $0.5$ brought non-convergence issues in our FEM simulations.
The mechanical tests are performed via a stationary approach and by imposing displacements.
For the compression tests (\figpref{fig1-1}{B,C}), the ring is compressed between two flat and infinitely rigid blocks, and contacts are solved via the Nitsche method.
For the interaction tests (\figref{fig4}), two facing rings are cut again along the $x$ direction to form semi-circles. 
To clamp the rings, no displacement in the $xz$ plane is imposed on the cut surfaces of the top ring.
The position of the bottom ring, first along $y$ for compression and then $x$ for shearing, is also controlled by imposing the displacement of its cut surfaces.
The contacts are also solved using the Nitsche method, but this time adding a Coulomb frictional interaction term with a unique friction coefficient $\mu$. 
The resulting shear force $f_x$ is then obtained by integrating the $\sigma_{yx}$ component of the stress on the surfaces at $y=0$ of the bottom ring, where the shear displacement is imposed.

\subsection*{Image acquisition and analysis}
Pictures are acquired via a CMOS camera (EMVC-CB1400C3 with an EMVL-MP814 lens, Misumi) placed on top of the assembly. Indirect lighting is obtained by shining a light-emitting diode panel on a whiteboard. 
Irrespective of shear amplitude, $15$ pictures per shear cycle are taken ($30$ pictures for $(\gamma_0,\Phi)=(0.400,0.94),(0.533,0.94),(0.533,1.00)$ where tracking is more challenging), and a $5\unit{s}$ pause between the cycles is implemented to ensure that each cycle's first and last pictures correspond to $\gamma=0$. 
The pictures consist of $4608\times1300\unit{pixels}$, resulting in a resolution of $13.05\unit{pixel/mm}$. Then, a custom-made Matlab script processes all the pictures. 
Using the orange-colored inner part of the rings, they are detected and separated via color and saturation thresholding. 
A group of pixels within its contour is attributed to each detected ring and, for instance, used to represent ring-specific quantities via color coding, as in \figpref{fig1-2}{A}. 
The geometrical center of the pixels defines then the position of the ring, while its shape, or contour, is measured via a circle Fourier decomposition \cite{saito_cell_2024}. 
The polar coordinates $R(\theta)$ of the outermost pixels of the contour are measured with the geometrical center of the ring as the origin. 
Using the fast Fourier transform, the Fourier coefficients $(a_k,b_k)$ of the contour are measured such that $R(\theta)=r_h\sum_{k=0}^M(a_k\cos\theta+b_k\sin\theta)$ up to $M=5$. 
In addition to the position and shape of the ring, we also detect their neighbors via two methods. 
First, we find topological neighbors via their position by performing the Delaunay tessellation.
Based on the resulting Voronoi diagram, we defined Voronoi neighbors as rings sharing a common edge and no farther than $2.5r$. 
From this Voronoi neighborhood, quantities such as the orientational bond order parameter $\Psi_6$ are computed (see \figref{fig2}). 
The other neighborhood detection method aims to detect the rings in contact. 
A given ring's contour is dilated by $8\unit{pixels}$, and its touching neighbors are those whose contour overlaps. 
For these touching rings, we then compute the bond vectors from their geometrical center and the contact extents for \figpref{fig4}{F}. 
To measure the extent of the contacts $\ell_c$, we compute the long axis of the overlapping contours. 
Indeed, each overlap has an approximate ellipsoidal shape with its long axis along the contacting surfaces (see ring overlaps in \supfigpref{S-fig_sup_schem}{B}). 
This measure overestimates the actual contact length because of the dilatation of the central ring, and we found that multiplying by $0.7$ gives a visually convincing measure.
Apart from $\Psi_6$, while we focused on the average (over all rings in the steady state) of the quantities of interest, their distribution is also available in \supfigref{S-fig_sup_distri}.

\subsection*{Geometrical and topological strains}
To distinguish quantitatively the geometrical deformations (change of shape) from the structural or topological ones (change in neighbors), we follow the method described in great detail in the Appendix section of Ref.\,\cite{guirao_unified_2015}. 
Initially developed for epithelial tissue development, this strain quantification can be simplified and applied to our ring assembly by taking into account the fact that ring assemblies are not confluent, and that the topological deformations arise only from neighbors' changes, in contrast to living cells which may also divide.
The geometric and topological strains, denoted respectively by $S^*$ and $T^*$, are computed between two consecutive time frames from the variations (length, orientation, existence) of the bonds between rings. 
The computation is based on the $2\times2$ texture tensors of the assembly, defined by $\mathbf{M}=\frac{1}{2}\sum_{B_\text{tot}}\Vec{L}\otimes\Vec{L}$ and $\mathbf{m}=\frac{1}{2}\sum_{b_\text{tot}}\Vec{l}\otimes\Vec{l}$, for the earlier and later consecutive frames, respectively. 
Here, $B_\text{tot}$ (resp.\,$b_\text{tot}$) is the set of all bond vectors $\Vec{L}$ (resp.\,$\Vec{l}$) in the assembly in the earlier (resp.\,later) frame, and $\Vec{L}\otimes\Vec{L} = \Vec{L}\Vec{L}^T$ is the tensor product of $\Vec{L}$ by itself.
Then we compute the deformation gradient tensor $\mathbf{F}=(\mathbf{F_0}\mathbf{F_1})^{1/2}$, with $\mathbf{F_0}=(\sum_{b_\text{cons}}\Vec{l}\otimes\Vec{L})(\sum_{b_\text{cons}}\Vec{L}\otimes\Vec{L})^{-1}$ and $\mathbf{F_1}=(\sum_{b_\text{cons}}\Vec{l}\otimes\Vec{l})(\sum_{b_\text{cons}}\Vec{L}\otimes\Vec{l})^{-1}$, where $b_\text{cons}$ is the set of the conserved bonds, i.e., the bonds that exist in both frames.
We also compute the strain correction tensor, $\mathbf{\Psi}=\mathbf{F}(\mathbf{M_c}\mathbf{F}^T-\mathbf{F}^T\mathbf{M_c})$, with $\mathbf{M_c}=\frac{1}{2}\sum_{b_\text{cons}}\Vec{L}\otimes\Vec{L}$.
For the contribution from the non-conserved bonds, we evaluate the topological tensor defined by $\mathbf{T}=\frac{1}{2}\sum_{b_\text{topo}}\Vec{l}\otimes\Vec{l}-\frac{1}{2}\sum_{B_\text{topo}}\Vec{L}\otimes\Vec{L}$, where $b_\text{topo}$ (resp.\,$B_\text{topo}$) is the set of the bonds that appear (resp.\,disappear) between the two frames (only due to rearrangements in our case, as opposed to living cells). 
Finally, we nondimensionalize each tensor $\mathbf{Q}$ by $\mathbf{\tilde{Q}}=\frac{1}{4}(\mathbf{Q}\mathbf{M_c}^{-1}+\mathbf{M_c}^{-1}\mathbf{Q})$ and obtain the desired strains by $\mathbf{S^*}=\frac{|B_\text{tot}|}{|b_\text{tot}|}\mathbf{\tilde{m}}-\mathbf{\tilde{M}}-\mathbf{\tilde{\Psi}}$ and $\mathbf{T^*}=\frac{|b_\text{tot}|-|B_\text{tot}|}{|b_\text{tot}|}\mathbf{\tilde{m}}-\mathbf{\tilde{T}}$, where $|B_\text{tot}|$ (resp.\,|$b_\text{tot}$|) is the number of the elements of the corresponding set, i.e., the total number of the bonds in the earlier (resp.\,later) frame.
Note that the total strain can be simply expressed, approximately, as $\mathbf{S^*} + \mathbf{T^*} \simeq \frac{1}{2}(\mathbf{F}\mathbf{F}^T-\mathbb{1})$ (valid for $\norm{\mathbf{S}^* + \mathbf{T}^*} \ll 1$). 
To sum up, by measuring the bonds in both frames and identifying the conserved, created and eliminated ones, we can compute the geometrical and topological strain tensors, $\mathbf{S^*}$ and $\mathbf{T^*}$, respectively.
To obtain simpler quantifiers of the geometrical and topological contributions, 
here we take the norm (the square root of the sum of the squared elements of the matrix) of these strain tensors and sum them over the $6$ shear cycles in the steady state.
The resulting quantities, $S_c=\int_t\norm{\mathbf{S^*}}\mathrm{d}t$ and $T_c=\int_t\norm{\mathbf{T^*}}\mathrm{d}t$, are the scalar quantities shown in \figpref{fig3}{E,F}.

\subsection*{Code and data availability}
Supplementary videos, experimental raw force data and data processed by image analysis of the ring assemblies, for all the parameters explored; FEM data and processed experimental data for the single ring compression and interactions; COMSOL files to run the FEM simulations and the associated Neo-Hookean material law; and raw images of the five first cycles for $\gamma_0=0.40$ and $\Phi = 1.00$, with a MATLAB code to process the images and extract the relevant quantities (each category comes with a dedicated README.txt that details its content) have been deposited in Zenodo \cite{zenodo_2024}. 
Some study data are available: due to size restrictions, all the experimental raw images of the ring assemblies are not included in the repository. 
However, the entire image dataset can be shared with interested researchers upon request.

}

\showmatmethods{} 

\acknow{This work benefited from numerous enriching discussions with, but not limited to, A. Ikeda, O. Dauchot, N. Saito, S. Ishihara, Y. Lou, H. Ikeda, F. Lechenault, C. Scalliet, M. Henot, I. Cantat, B. Chakraborty, and B. Andreotti. We thank R. Tosaka for the construction of a prototype of the shear device. This work is supported in part by KAKENHI from the Japan Society for the Promotion of Science (Grant Nos. JP22KF0084, JP20H00128, JP19H05800, and JP24K00593). S.P. acknowledges financial support as a Japan Society for the Promotion of Science International Research Fellow.}

\showacknow{} 


\bibliography{ref}

\end{document}



\maketitle



\begin{figure}
\includegraphics[]{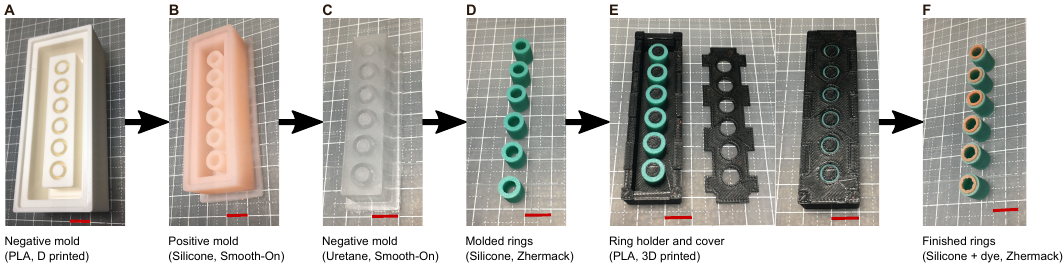}
\centering
\caption{Fabrication process of the rings. 
A negative mold 3D printed with PLA (A) is used to fabricate a positive mold made of silicone (B). The silicone mold itself is used to fabricate a urethane negative mold (C), in which the rings are made (D). Then, the rings are placed in a 3D printed holder with a cover (E), leaving only the internal half of the ring top surface visible. A thin layer of dyed silicone is then coated on the surface to give the final rings (F). Before experiments, the rings are placed in a sealed plastic bag with talc powder and thoroughly shaken to ensure a homogeneous coating of the rings. The multiple molds were used to satisfy the following constraints: the mold needs to be flexible to ensure demolding without breaking the rings, silicone sticks to silicone molds, but not at all with urethane (also a soft polymer), urethane sticks to everything except silicone. Only the choice of the silicone for the rings and its colored top is crucial (Elite Double 22, Zhermack) to ensure no plastic deformation of the rings, even under prolonged and high deformations. The red scale bar in the pictures measures $10\unit{mm}$.
}
\label{fig_sup_ring}
\end{figure}

\begin{table}
\centering
\begin{tabular}{|*{5}{m{3cm}|}}

\hline
\cellcolor[RGB]{141,160,203} $\gamma_0=0.667$\newline$(X_0=50\unit{mm})$ & \cellcolor[RGB]{240,249,232} $N_c=100$\newline$R_\mathrm{tracked}=84.8\%$ & \cellcolor[RGB]{186,228,188} $N_c=15$\newline$R_\mathrm{tracked}=91.2\%$ & \cellcolor[RGB]{240,249,232} $N_c=15$\newline$R_\mathrm{tracked}=95.9\%$ & \cellcolor[RGB]{186,228,188} $N_c=15$\newline$R_\mathrm{tracked}=96.7\%$\\[10pt]
\hline
\cellcolor[RGB]{141,160,203} $\gamma_0=0.533$\newline$(X_0=40\unit{mm})$ & \cellcolor[RGB]{240,249,232} $N_c=100$\newline$R_\mathrm{tracked}=92.9\%$ & \cellcolor[RGB]{186,228,188} $N_c=50$\newline$R_\mathrm{tracked}=95.9\%$ & \cellcolor[RGB]{240,249,232} $N_c=40$\newline$R_\mathrm{tracked}=97.4\%$ & \cellcolor[RGB]{186,228,188} $N_c=45$\newline$R_\mathrm{tracked}=97.5\%$\\[10pt]
\hline
\cellcolor[RGB]{141,160,203} $\gamma_0=0.400$\newline$(X_0=30\unit{mm})$ & \cellcolor[RGB]{240,249,232} $N_c=100$\newline$R_\mathrm{tracked}=98.3\%$ & \cellcolor[RGB]{186,228,188} $N_c=100$\newline$R_\mathrm{tracked}=98.7\%$ & \cellcolor[RGB]{240,249,232} $N_c=60$\newline$R_\mathrm{tracked}=99.1\%$ & \cellcolor[RGB]{186,228,188} $N_c=75$\newline$R_\mathrm{tracked}=95.3\%$\\[10pt]
\hline
\cellcolor[RGB]{141,160,203} $\gamma_0=0.267$\newline$(X_0=20\unit{mm})$ & \cellcolor[RGB]{240,249,232} $N_c=100$\newline$R_\mathrm{tracked}=99.3\%$ & \cellcolor[RGB]{186,228,188} $N_c=100$\newline$R_\mathrm{tracked}=98.4\%$ & \cellcolor[RGB]{240,249,232} $N_c=100$\newline$R_\mathrm{tracked}=100\%$ & \cellcolor[RGB]{186,228,188} $N_c=100$\newline$R_\mathrm{tracked}=99.7\%$\\[10pt]
\hline
\cellcolor[RGB]{141,160,203} $\gamma_0=0.133$\newline$(X_0=10\unit{mm})$ & \cellcolor[RGB]{240,249,232} $N_c=100$\newline$R_\mathrm{tracked}=100\%$ & \cellcolor[RGB]{186,228,188} $N_c=100$\newline$R_\mathrm{tracked}=100\%$ & \cellcolor[RGB]{240,249,232} $N_c=100$\newline$R_\mathrm{tracked}=99.4\%$ & \cellcolor[RGB]{186,228,188} $N_c=100$\newline$R_\mathrm{tracked}=99.2\%$\\[10pt]
\hline
\cellcolor[RGB]{141,160,203} $\gamma_0=0.067$\newline$(X_0=5\unit{mm})$ & \cellcolor[RGB]{240,249,232} $N_c=100$\newline$R_\mathrm{tracked}=99.7\%$ & \cellcolor[RGB]{186,228,188} $N_c=100$\newline$R_\mathrm{tracked}=99.7\%$ & \cellcolor[RGB]{240,249,232} $N_c=100$\newline$R_\mathrm{tracked}=100\%$ & \cellcolor[RGB]{186,228,188} $N_c=100$\newline$R_\mathrm{tracked}=100\%$\\[10pt]
\hline
 & \cellcolor[RGB]{141,160,203} $\Phi=0.94$\newline$(N=296)$ & \cellcolor[RGB]{141,160,203} $\Phi=1.00$\newline$(N=320)$ & \cellcolor[RGB]{141,160,203} $\Phi=1.06$\newline$(N=340)$ & \cellcolor[RGB]{141,160,203} $\Phi=1.12$\newline$(N=360)$ \\[10pt]
 \hline
\end{tabular}
\vspace{3mm}
\caption{Experimental parameters for the shear experiments. The shear amplitude $\gamma_0$ is set by the lateral displacement amplitude of the top arm, $X_0$, while the density $\Phi$ is controlled by the number of rings $N$. For each set of the parameters $(\gamma_0,\Phi)$, the total number of cycles $N_c$ (including the transient and the steady state) and the proportion of rings successfully tracked $R_\mathrm{tracked}$ are indicated. 
The last $6$ available cycles (for instance, from cycle $45$ to $50$ if $N_c=50$) are considered to be in the steady state and are used for the analyses presented in this work.
Progressive out-of-plane extrusion of some rings limits the number of valid cycles for high shear and density. Due to the extreme deformability of the rings, upon cycling, some rings manage to escape the plane of the assembly despite the cover plate by out-of-plane deformations. In \figref{fig_sup_same_cycle}, we verify that the results and following conclusions are not dependent on which cycles we consider as the steady state and make the measurements.}
\label{suptab1}
\end{table}

\begin{figure*}
\includegraphics[width=\hsize,clip]{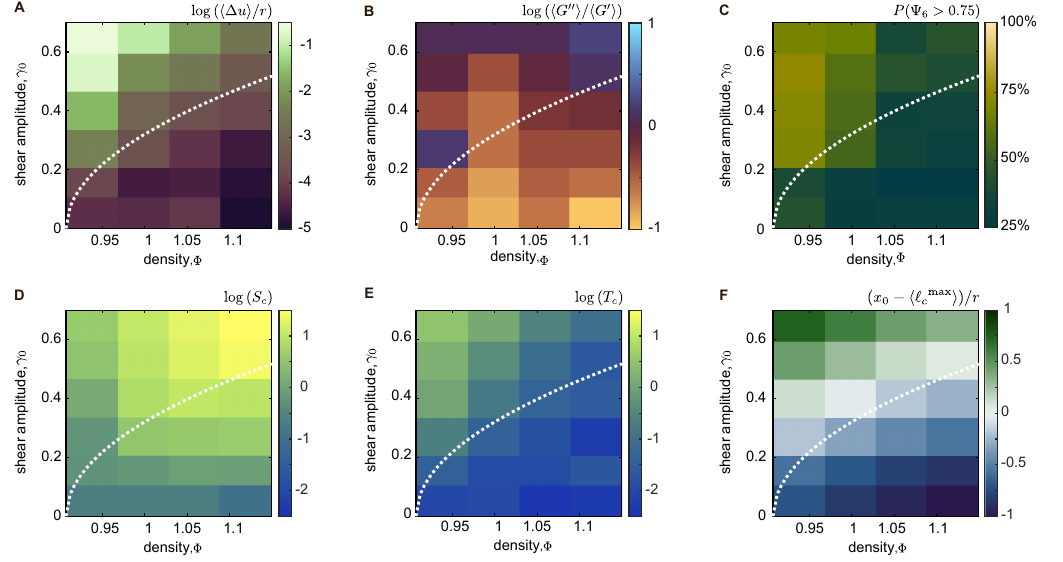}
\centering
\caption{The quantities in the $(\gamma_0,\Phi)$ parameter space are not sensitive to the definition of the steady state.
The panels shown here correspond to the parameter maps presented in the main text, but obtained for the cycle numbers fixed to $10$ to $15$ for all the parameters, the highest numbers common to all parameters. Note that, for these cycle numbers, the steady state was not reached for some parameters. 
(A) Irreversible displacement. (B) Ratio between the viscous and elastic moduli. (C) Proportion of rings in a crystal cluster. (D) Geometrical strain. (E) Topological strain. (F) Difference between the imposed lateral displacement and maximum contact length. Because the steady state was not fully reached in the data shown here, some quantitative differences can be observed in the proportion of crystal (C) and shear moduli (B), but the tendency and evolution remain essentially the same. The choice of the cycles to take into account is, then, not a determining factor in analyzing the response of ring assemblies.
}
\label{fig_sup_same_cycle}
\end{figure*}

\begin{figure}[]
    \centering
    \includegraphics[]{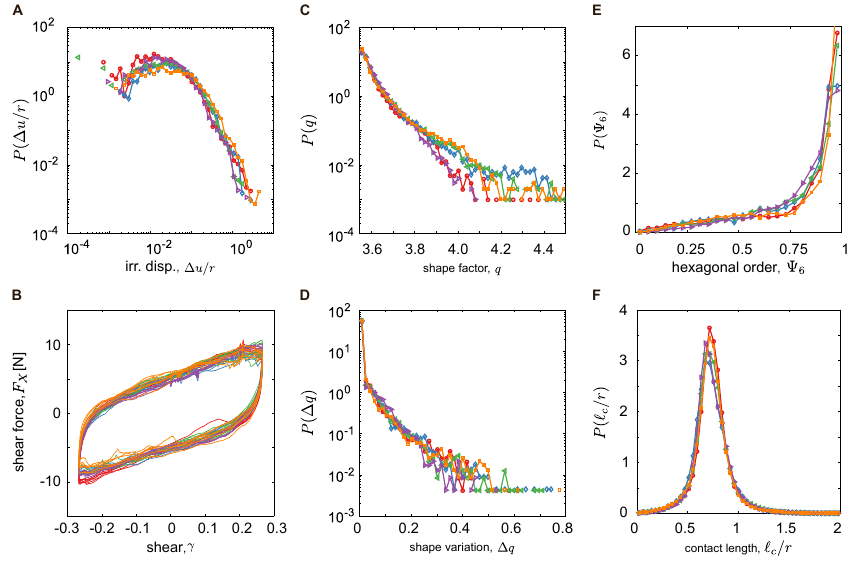}
    \caption{Reproducibility of the results, tested for $(\gamma_0,\Phi)=(0.4,1.00)$ by repeating the same experiments $5$ times with different initial configurations. The probability distributions of relevant quantities and the force response curves for the $5$ experiments are plotted by different colors and symbols: (A) irreversible displacement $\Delta u /r$, (B) shear force $F_X$, (C) shape factor $q$, (D) shape variation $\Delta q$, (E) hexagonal order $\Psi_6$, and (F) contact length $\ell_c/r$. The absence of significant differences in the analyzed quantities indicates that our experiments are highly reproducible and not sensitive to the initial state.
    }
    \label{fig_sup_repro}
\end{figure}

\begin{figure}[]
\centering
\includegraphics[]{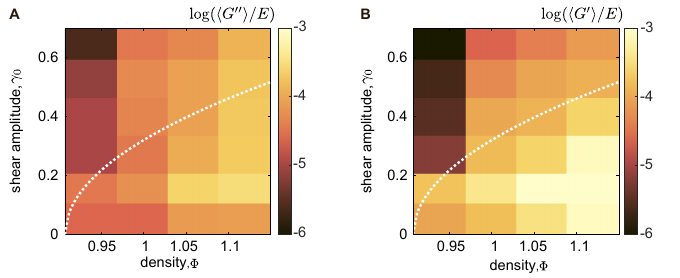}
\caption{Loss and storage moduli across the parameter space.
The first harmonic of the loss and storage moduli are shown in (A) and (B), respectively. The moduli are normalized by the Young modulus of the 
rings, $E=1\unit{MPa}$. Details on how the force response and moduli are measured can be found in the Materials and Methods section "Force measurement".}
\label{fig_sup_moduli}
\end{figure}

\begin{figure}[]
\centering
\includegraphics[]{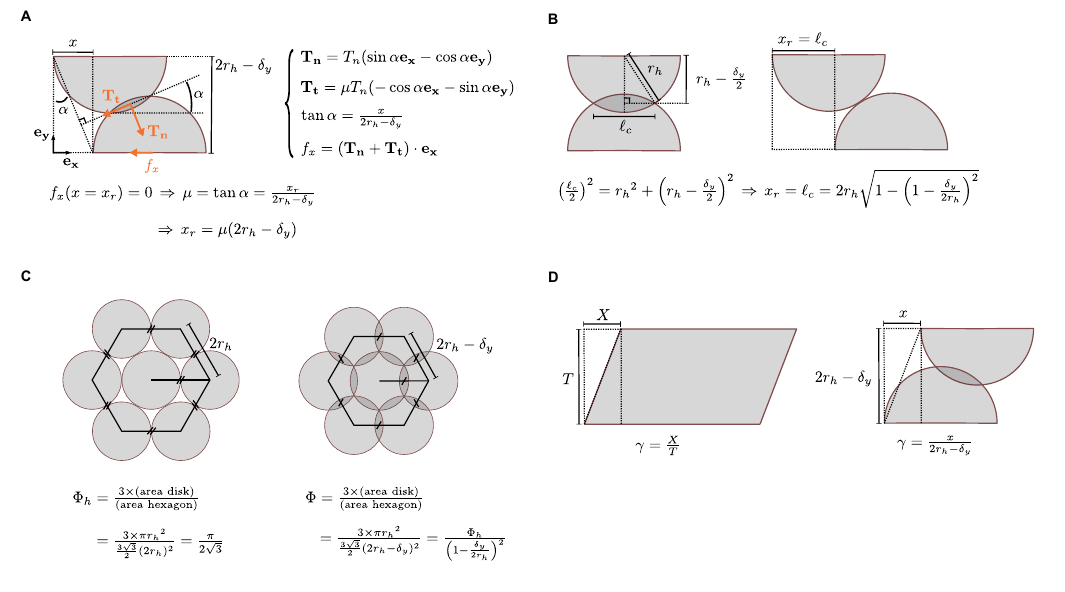}
\caption{Schematic for the derivation of the geometrical predictions. 
(A) Estimation of the lateral displacement $x_r$ for the first case, in which the lateral force $f_x$ becomes repulsive ($f_x>0$) before the two rings fully separate. The contact is reduced to a single point with an orientation $\alpha$ given by the line linking the two intersection points of the half-circles. Assuming frictional Coulomb interactions, the amplitude of tangential $\mathbf{T_t}$ and normal $\mathbf{T_n}$  contact forces are linked by the friction coefficient $\mu$. The lateral force $f_x$ is the resultant of the contact forces projected on the lateral axis: $f_x=(\mathbf{T_n}+\mathbf{T_t})\cdot\mathbf{e_x}$. The expression for $x_r$ is then found by solving the equation $f_x(x=x_r)=0$. This prediction is valid if the two rings are still in contact at $x=x_r=\mu(2r_r-\delta_y)$. 
(B) If the rings are not in contact at $x=x_r=\mu(2r_r-\delta_y)$, we have to consider the second case where $f_x$ remains attractive until complete separation of the rings. In that case, $x_r$ is the no-overlap distance given by the initial contact extent $\ell_c$ after compression. 
(C) To link the global assembly density $\Phi$ and the local compression $\delta_y$, we assume an isotropic and homogeneous compression of an assembly initially in a hexagonal packing. The density is estimated by considering a central ring and the hexagon with corners at the center of its $6$ neighbors and by taking the ratio between the area of the ring inside the hexagon ($1$ full ring + $6$ third of rings) and the area of the hexagon. In the hexagonal lattice, the hexagon edge is $2r_h$ and reduces to $2r_h-\delta_y$ under compression, while the area of the rings is constant ($\Phi$ is estimated from the area of undeformed rings).
(D) Shear is a scaleless quantity defined as the ratio between the lateral displacement and the height of the system. The shear defined for the assembly as $\gamma_=X/T$ is then identical to the one in the local interaction experiments, $\gamma=x/(2r_h-\delta_y)$. 
}
\label{fig_sup_schem}
\end{figure}

\begin{figure}[]
\centering
\includegraphics[]{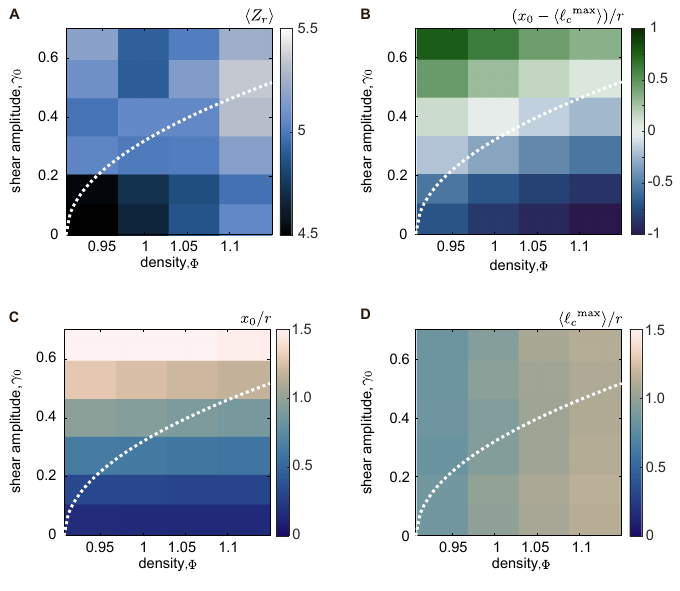}
\caption{The average coordination number is irrelevant for highly compressed assemblies.
(A) Average of the coordination number $Z_r$ of the assemblies in the steady state. The coordination number of a ring is defined as the number of touching neighbors. No clear tendency with $\gamma_0$, $\Phi$, or the yielded/solid regions is noticeable.
(B) Reproduction of \figpref{M-fig4}{F}, showing that the relative value between the shear distance $x_0$ and the averaged maximal contact length $\langle{\ell_{c}}^{\mathrm{max}}\rangle$ marks a clear separation between yielded and solid-like regions.
(C) Local average lateral displacement $x_0$ imposed to the assembly over the parameter space $(\gamma_0,\Phi)$. It is a deterministic quantity computed following $x_0=\gamma_02r_h\sqrt{\Phi_h/\Phi}$.
(D) For each tracked ring in the steady state, the contact length $\ell_c$ with each of its touching neighbors is measured (see Methods). $\langle{\ell_{c}}^{\mathrm{max}}\rangle$ is the average of the highest $\ell_c$ of each ring over all rings and all time steps in the steady state. 
While $x_0$ mainly increases with the shear $\gamma_0$, $\langle{\ell_{c}}^{\mathrm{max}}\rangle$ is increasing with density $\Phi$. The two values coincide around the yielding separation line predicted by $\gamma_r(\Phi)=\sqrt{\Phi/\Phi_h-1}$.
}
\label{fig_sup_coor}
\end{figure}

\begin{figure}[]
    \centering
    \includegraphics[]{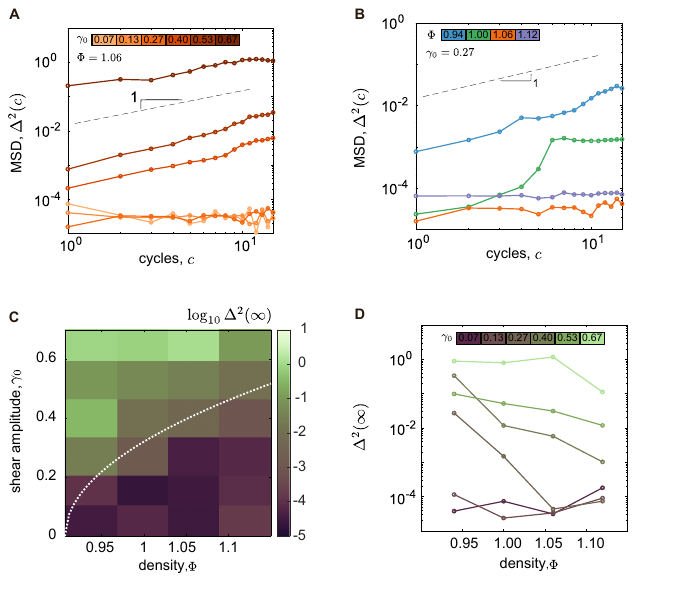}
    \caption{
    Analysis of the mean-squared displacement (MSD) of the rings. 
    For each parameter pair $(\gamma_0,\Phi)$, we computed the dimensionless MSD defined as $\Delta^2(c)=\frac{1}{N}\sum\limits_{i=1}^N\left({\frac{u(i,c)-u_0(i)}{2r}}\right)^2$, with $u(i,c)$ the position of ring $i$ at the start of cycle $c$, $u_0(i)$ its position at the reference cycle, $N$ the number of tracked rings, and $r$ their radius. The reference cycle was chosen to be $15$ cycles before the final cycle (see Table\,\ref{suptab1} for the total number of cycles, $N_c$). 
    (A) MSD for a fixed density $\Phi=1.06$ and varying shear amplitudes $\gamma_0$. We observe a transition from almost immobile rings in the solid region (small $\gamma_0$) to diffusive-like motion of rings in the yielded region (large $\gamma_0$). 
    (B) MSD for a fixed shear amplitude $\gamma_0=0.27$ and varying densities $\Phi$. As the density decreases and the assembly enters the yielded region, the trajectories of the rings become diffusive. 
    (C) The Debye-Waller (DW) $\Delta^2(\infty)$ factor, defined as the long-time limit of the MSD. With our limited dataset, we compute the DW factor as the average of the MSD over the last three cycles. 
    (D) The DW factor as a function of density $\Phi$ for different shear amplitudes $\gamma_0$. As elucidated in Ref.\,[56], a progressive decrease of the DW factor for large $\gamma_0$ is a characteristic of a glass transition with dominating thermal-like effects, while an abrupt drop observed for $\gamma_0=0.27$ hints at an underlying jamming transition. For $\gamma_0<0.27$, likely, the limited number of cycles we could carry out did not allow us to observe a signature of the transition. 
    }
    \label{fig_sup_transition}
\end{figure}

\begin{figure}[]
    \centering
    \includegraphics[]{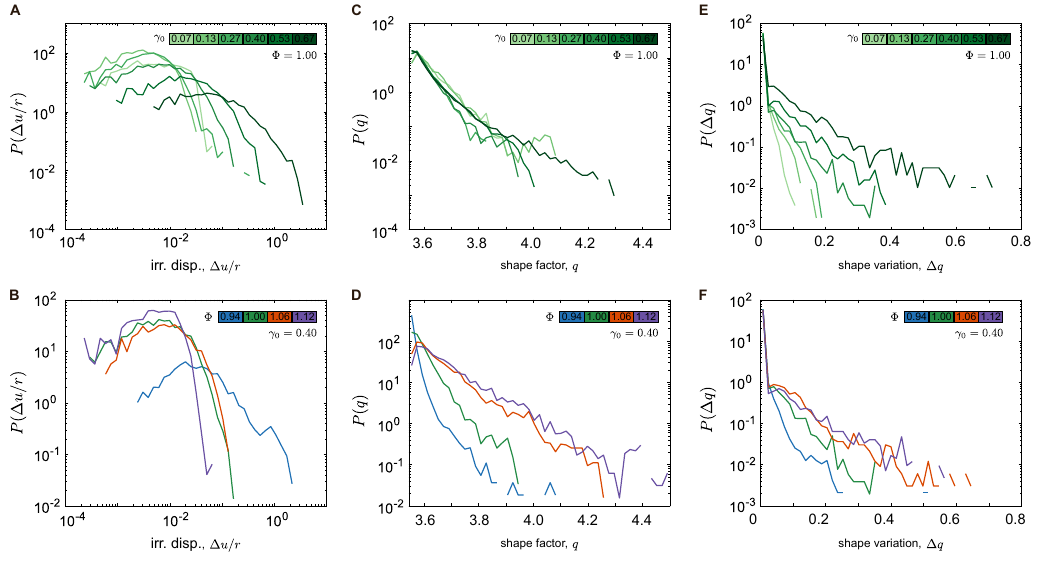}
    \caption{Probability distributions of representative quantities, taken over all rings in the steady states. Top, distributions for $\Phi=1.00$ and different $\gamma_0$; bottom, distributions for $\gamma_0$ and different $\Phi$. 
    (A,B): irreversible displacement between two consecutive cycles normalized by the rings' radius.
    (C,D): shape factor $q$ over all time steps.
    (E,F): shape change over a cycle $\Delta q$.
    }
    \label{fig_sup_distri}
\end{figure}
    
\FloatBarrier

\section*{Movie captions}

\movie{
Video of the first ten shear cycles for the assemblies with $(\gamma_0,\Phi)=(0.4,0.94)$ (top) and $(\gamma_0,\Phi)=(0.4,1.16)$ (bottom).}

\movie{
Illustration of the measurements made on the rings, here shown for a single ring and parameters $(\gamma_0,\Phi)=(0.4,1.0)$. The shape is highlighted in red, and the Voronoi diagram is in fine white dotted lines. Each touching ring and its corresponding bond vector is white, while the contact width is blue.}

\movie{
Non-affine trajectories of the rings in the steady state. The affine displacement due to the instantaneous shear strain is retrieved for each ring trajectory and time frame along shear cycles.}

\movie{
An animated version of \figpref{M-fig2}{A,B} showing the probability distribution of $\Psi_6$ with the number of shear cycles.}

\movie{
Stroboscopic evolution of the structure of the assemblies with the number of shear cycles. The video shows, for all the experiments, snapshots of the assembly at the start of each cycle with each tracked ring colored by its $\Psi_6$ value. The assembly is darkened when it reaches its maximal cycling number.}

\movie{
Bond vector dynamics of the assemblies. The video shows, for all the experiments, snapshots of the assembly upon cycling in the steady state. For each picture, the links that are conserved are colored in white, the ones disappearing at the next frame in blue, and the ones appearing from the previous frame in red.}

\movie{
Video showing the FEM simulation of ring-ring experiments, with snapshots of the rings on the left, and the corresponding shear force $f_x(x)$ in the right. The results for $\mu=0.3$ and $\mu=0.6$ are shown for a compression of $\delta_y=950\unit{\mu m}$.}
